%% file: main.tex
\documentclass[preprint,12pt]{elsarticle}

\usepackage[T1]{fontenc}
\usepackage[utf8]{inputenc}
\usepackage{lmodern}
\usepackage{textcomp}
\usepackage{amsmath,amssymb}
\usepackage{graphicx}
\usepackage{booktabs}
\usepackage{tabularx}
\usepackage{array}
\usepackage{microtype}
\usepackage[section]{placeins} 
\usepackage{float}
\newfloat{listing}{tbp}{lol}
\floatname{listing}{Listing}
\usepackage{url}
\usepackage[hidelinks]{hyperref}
\usepackage{lineno}

\graphicspath{{figures/}}

\newcolumntype{L}{>{\raggedright\arraybackslash}X}
\input{macros}

\makeatletter
\def\ps@pprintTitle{\let\@oddhead\@empty\let\@evenhead\@empty
  \def\@oddfoot{\reset@font\hfil\thepage\hfil}\let\@evenfoot\@oddfoot}
\makeatother
\biboptions{sort&compress}

\begin{document}

\begin{frontmatter}

\title{TARANG: an open-source, cost-controlled design workflow for silicon nitride microcomb chips}

\author[loqm]{Divyanshu Tiwari}
\author[plaksha]{Esraaj Sarkar Gupta}
\author[loqm]{Anshuman Kumar\corref{cor1}}
\ead{anshuman.kumar@iitb.ac.in}
\cortext[cor1]{Corresponding author}
\affiliation[loqm]{organization={Laboratory of Optics of Quantum Materials, Department of Physics, Indian Institute of Technology Bombay},
            city={Mumbai},
            postcode={400076},
            country={India}}
\affiliation[plaksha]{organization={Plaksha University},
            city={Mohali},
            state={Punjab},
            country={India}}

\begin{abstract}
Designing a microresonator frequency comb chip means solving a chain of coupled problems: a waveguide with anomalous dispersion that survives fabrication errors, a ring with the right free spectral range and quality factor, a bus gap that sets the coupling regime, and grating couplers, all made in one lithography step. We present TARANG, an open-source Python package that takes an air clad silicon nitride platform from tabulated refractive indices to a GDSII mask in fourteen stages, built on the Tidy3D mode, FDTD and eigenmode-expansion solvers. Every cloud simulation is priced before it runs and capped, every result is cached by a fingerprint of the full simulation, and every quoted number carries a numerical check. We show that snapping the grid onto the core's faces, a common way to steady dispersion calculations, leaves the field samples on the faces to floating-point rounding, and that averaging solves with the grid just inside and just outside the faces makes the group-velocity dispersion $D$ independent of the computer and steady to about 1\Dunit{} from mesh to mesh. Two complete designs are presented, at 1550\,nm ($700\times1000$\,nm core, $D = 372.9 \pm 0.6$\Dunit) and at 780\,nm ($350\times500$\,nm, $D = 196.4 \pm 0.9$\Dunit), each with a 200\,GHz ring, critical and over coupled gaps, grating couplers, and a fiber to fiber budget. The predicted dispersion agrees with two measured devices from the literature to within 1 and 4\pct, and an independent Lugiato--Lefever solver forms stable single solitons from the exported parameters of both designs, which are ready for fabrication.
\end{abstract}

\begin{keyword}
Silicon nitride \sep Microresonator frequency combs \sep Dispersion engineering \sep Grating couplers \sep FDTD \sep Open source design software
\end{keyword}

\end{frontmatter}

\input{sections/00_program_summary}
\clearpage

\input{sections/01_introduction}

\input{sections/02_workflow}

\input{sections/03_waveguide}

\input{sections/04_ring}

\input{sections/05_gratings}

\input{sections/06_comparison}

\input{sections/07_validation}

\input{sections/08_conclusions}

\input{sections/09_back_matter}

\bibliographystyle{elsarticle-num}
\bibliography{refs}

\end{document}


\begin{frontmatter}
\title{Supplementary Material for\\ ``TARANG: an open-source, cost-controlled design workflow for silicon nitride microcomb chips''}
\author[loqm]{Divyanshu Tiwari}
\author[plaksha]{Esraaj Sarkar Gupta}
\author[loqm]{Anshuman Kumar}
\affiliation[loqm]{organization={Laboratory of Optics of Quantum Materials, Department of Physics, Indian Institute of Technology Bombay},
            city={Mumbai}, postcode={400076}, country={India}}
\affiliation[plaksha]{organization={Plaksha University}, city={Mohali}, state={Punjab}, country={India}}
\end{frontmatter}

\noindent This document gives the details and further checks behind the main paper. Sections, figures, tables and equations here are numbered S1, S2, \ldots; numbers without an S refer to the main paper. All results were computed with TARANG v0.1.5 and Tidy3D 2.12.0.

\input{sections_supplementary/S1_materials}

\input{sections_supplementary/S2_cross_section}

\input{sections_supplementary/S3_ring_bus_coupling}

\input{sections_supplementary/S4_grating_numerics}

\input{sections_supplementary/S5_grating_tolerances}

\input{sections_supplementary/S6_whole_device}

\input{sections_supplementary/S7_solitons}

\input{sections_supplementary/S8_sloped_walls}

\input{sections_supplementary/S9_reproducing}

\bibliographystyle{elsarticle-num}
\bibliography{refs}

%% file: macros.tex
\newcommand{\um}{\,\textmu m}
\newcommand{\Dunit}{\,ps\,nm$^{-1}$\,km$^{-1}$}
\newcommand{\TE}{TE$_{00}$}
\newcommand{\TM}{TM$_{00}$}
\newcommand{\SiN}{Si$_3$N$_4$}
\newcommand{\SiO}{SiO$_2$}
\newcommand{\pct}{\,\%}
\newcommand{\neff}{n_\mathrm{eff}}
\newcommand{\Dint}{D_\mathrm{int}}

%% file: sections/00_program_summary.tex
\section*{Program summary}
\noindent\textit{Program title:} TARANG\\
\textit{CPC Library link to program files:} (to be added by Technical Editor)\\
\textit{Developer's repository link:} \url{https://github.com/loqm/tarang}\\
\textit{Licensing provisions:} MIT\\
\textit{Programming language:} Python ($\geq$ 3.11)\\
\textit{External routines/libraries:} Tidy3D 2.12~\cite{tidy3d} (LGPL-2.1-or-later; its cloud solvers need a Flexcompute account), NumPy~\cite{numpy2020}, Matplotlib~\cite{matplotlib2007}, gdstk~\cite{gdstk} (layout); pyLLE~\cite{moille2019} (optional, soliton simulations)\\
\textit{Nature of problem:} Designing an air-clad, fully etched \SiN{} microresonator frequency-comb chip couples several problems that are usually solved separately and by hand: the waveguide dispersion and its fabrication tolerance, the ring radius and quality factor, the ring--bus coupling, and grating couplers made in the same full etch. The three-dimensional simulations involved run on a paid cloud solver, so their cost has to be controlled.\\
\textit{Solution method:} A fourteen-stage workflow built on the Tidy3D finite-difference eigenmode, FDTD and eigenmode-expansion solvers: dispersion maps with two-sided grid snapping and a mesh ladder; the ring radius from the free spectral range; 3D FDTD of the ring--bus coupler, cross-checked with coupled-mode theory; 2D FDTD design of uniform, apodised and focusing grating couplers; tolerance and sidewall sweeps; a whole-device fibre-to-fibre spectrum; a GDSII layout; and export of the Lugiato--Lefever inputs for pyLLE. Every cloud run is priced before it is submitted and capped, and every result is cached by a fingerprint of the full simulation.\\
\textit{Additional comments including restrictions and unusual features:} Cloud FDTD and EME runs need a Tidy3D account and FlexCredits; local mode solving is free. A design folder records every setting, result, figure with its data, assumed value with its source, and software version.

%% file: sections/01_introduction.tex
\section{Introduction}\label{sec:intro}

A dissipative Kerr soliton in a high $Q$ microresonator turns a continuous wave pump into a broad, phase coherent frequency comb on a chip~\cite{kippenberg2018,gaeta2019,herr2014}. Its physics is described by the Lugiato--Lefever equation~\cite{lugiato1987,chembo2013,coen2013}, whose inputs (the ring's integrated dispersion, its intrinsic and coupling losses, and its effective nonlinearity) are the quantities a chip designer can control. Bright solitons need anomalous group velocity dispersion at the pump, which in a waveguide means overcoming the material's normal dispersion with the geometric dispersion of a strongly confined mode, which confines the core height and width to narrow windows. Stoichiometric silicon nitride is the standard material~\cite{moss2013}, and an air clad, fully etched \SiN{} core on thermal oxide is the simplest platform, patterned by electron-beam lithography in university cleanrooms, down to visible wavelengths~\cite{mandal2024emission,mandal2026stitchfree}.

The design of such a chip is a chain. The cross section sets the dispersion and the group index; the group index and the free spectral range (FSR) set the ring radius; the radius, the losses and the cross-section set the gap for critical coupling; the film and the etch constrain the gratings; and fabrication errors in width, thickness and sidewall angle move every link. Each link is usually solved by hand with separate scripts, which makes designs hard to reproduce and check, and cloud solvers add a cost to every iteration.

Many tools solve single links of this chain. Commercial suites such as Ansys Lumerical and cloud solvers such as Tidy3D~\cite{tidy3d} provide mode, FDTD and eigenmode-expansion solvers; open-source codes such as MPB~\cite{mpb2001}, Meep~\cite{meep2010} and femwell~\cite{femwell} provide mode and FDTD solvers; gdsfactory~\cite{gdsfactory} generates mask layouts and connects them to several of these simulators; and Lugiato--Lefever solvers such as pyLLE~\cite{moille2019} simulate the comb. The decisions between them---which cross-section is robust to fabrication, which radius and gap follow from it, which grating the film allows, how fabrication errors propagate---and the checks that each result is converged are left to the user, usually in scripts written for one design. Within the comb community, recent work optimises the dispersion itself. Genetic algorithm~\cite{lucas2023} and adjoint~\cite{chuchalin2026} inverse design find the integrated dispersion that gives a target comb spectrum, and machine learning infers the ring geometry from a dispersion profile~\cite{pal2023}. These methods address the dispersion step; they do not carry the design to the couplers, the layout and the fabrication tolerances, which is where TARANG starts from a target dispersion and ends at a mask.

TARANG automates this chain. Rather than adding a new solver, it builds on the Tidy3D eigenmode, FDTD and eigenmode-expansion solvers~\cite{tidy3d} and exports the inputs of the Lugiato--Lefever solver pyLLE~\cite{moille2019}. It adds the design logic between the solvers, physics checks at each stage, cost control for cloud simulation, and a design folder that records every number with its source. This paper describes the workflow (Section~\ref{sec:workflow}), the physics and numerics of each stage (Sections~\ref{sec:waveguide}--\ref{sec:grating}), two complete designs at 1550 and 780\,nm (Section~\ref{sec:compare}), and their validation (Section~\ref{sec:validation}). Snapping the grid onto a core's faces leaves the field samples on the faces to floating-point rounding, and a two-sided average removes the effect (Section~\ref{sec:waveguide}). Details, further checks and the data behind every figure are in the Supplementary Material.

%% file: sections/02_workflow.tex
\section{Workflow and software}\label{sec:workflow}

\begin{figure}[htbp]
\centering
\includegraphics[width=\textwidth]{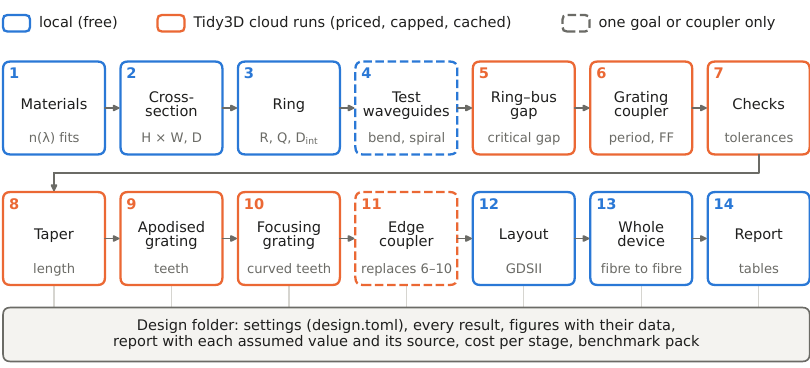}
\caption{The TARANG workflow. Fourteen stages run in order, each reading the results of the stages before it. Blue: local mode solves (free). Orange: stages that use Tidy3D cloud simulations (stage 10 only for its optional 3D check), each priced before it is submitted, capped per run and per design, and cached by a fingerprint of the full simulation. Dashed: stages used only for one goal (test waveguides) or one coupler type (edge couplers instead of stages 6--10). Stage 2 maps the dispersion, picks the robust target and checks the mesh; stage 7 checks tolerances, mode families, losses, the sidewall angle and the grating budget; stage 12 also exports the parameters for pyLLE. Every stage writes into the design folder.}\label{fig:workflow}
\end{figure}

TARANG designs microcomb chips on a planar stack: a fully etched core on a buried oxide over a silicon handle, under air or a top cladding. Each layer's material is a table of refractive indices, built in or the user's own, and the thicknesses of the film, the oxide and the cladding are settings. The defaults, and both designs in this paper, are air-clad: a stoichiometric \SiN{} core on 4\um{} of thermal oxide (\SiO{}) with no top cladding, the stack used to reach anomalous dispersion at visible wavelengths~\cite{moille2025}. A design is a run of fourteen stages
(Fig.~\ref{fig:workflow}). Each stage reads the results of the stages before it, takes its settings from a \texttt{design.toml} file or a browser interface, and writes its results, figures and tables to a design folder. Cross-sections are mode-solved locally, at no cost; three-dimensional FDTD and eigenmode-expansion (EME) runs (in stages 5--11) run on the Tidy3D cloud and cost FlexCredits. A design is set by a short settings file (Listing~\ref{lst:toml}). The 1550\,nm design used 163 cloud simulations and 12.0 FlexCredits. The 780\,nm design used 170 simulations and 62.6 FlexCredits, mostly for its finer grating mesh and its 3D coupler runs, which are larger in wavelengths (Supplementary Table~S10).

\begin{listing}[H]
\scriptsize
\begin{minipage}[t]{0.47\textwidth}
\begin{verbatim}
[run]
goal = "comb_ring"
coupling = "grating"
wavelength_um = 1.55
max_cost_per_run = 6.0  # FlexCredits
max_total_cost = 40.0
[ring]
fsr_ghz = 200.0
loss_preset = "typical_ebl"  # 0.4 dB/cm
\end{verbatim}
\end{minipage}\hfill
\begin{minipage}[t]{0.47\textwidth}
\begin{verbatim}
[cross_section]
heights_um = [0.68, 0.7, 0.72]
widths_um = [0.6, 0.7, 0.8, 0.9,
  1.0, 1.1, 1.2, 1.3, 1.4,
  1.5, 1.6, 1.7, 1.8]
mesh_ladder = [50, 60, 70, 80, 90]
\end{verbatim}
\end{minipage}
\caption{Settings of the 1550\,nm design (excerpt of \texttt{design.toml}; full file in Supplementary Listing~S1), run with \texttt{python -m tarang.design -{}-config design.toml}. Settings not given take documented defaults.}\label{lst:toml}
\end{listing}

Four features carry the claims of reproducibility and cost control. (i)~\emph{Priced and capped cloud runs.} The cost of each simulation is estimated before submission, and it runs only within a per-run and a per-design budget set by the user; a \texttt{-{}-price-only} mode prices a whole design without spending anything. (ii)~\emph{A content-addressed cache.} Each cloud result is stored under a fingerprint of the full simulation description and reused only for an identical simulation, so repeating a design costs nothing and an interrupted run resumes where it stopped. (iii)~\emph{Failures are reported, never used.} A simulation whose fields diverged is reported with a link to the solver's record and its result is not used: the mesh and boundary checks list such a run, or an unreadable one, and leave it out of their tables and figures, and in the coupler and grating stages it stops the stage. (iv)~\emph{The design folder is the record.} It holds the settings, every result in machine-readable form, each figure with its data, a report that lists every assumed value with its published source, the software versions, the cost of each stage, and a benchmark pack for repeating the key simulations with another solver. The layout stage adds GDSII cells and the parameters of every device (JSON, CSV and MATLAB) for other layout tools.

\paragraph{Scope and limitations} TARANG designs ring resonators with their bus, couplers and test waveguides on the stack described above. The gratings can be fully or partially etched or replaced by edge couplers, but the defaults and the validation in this paper cover air-clad, fully etched \SiN. Gratings are designed in 2D FDTD, with a 3D check of the focusing grating as an option. The material data are bulk values and the losses are assumed values with published sources; these are the largest systematic errors before a chip is measured, and both can be replaced by measured data. The whole device model leaves out back scattering in the ring, the phase of the coupler and the reflection of the taper. Thermal effects enter only as the static shift of the resonances per kelvin; the thermal dynamics that govern access to soliton states are not modelled, and the comb itself is simulated by the Lugiato--Lefever solver to which TARANG exports its results.

%% file: sections/03_waveguide.tex
\section{Waveguide dispersion}\label{sec:waveguide}

\begin{figure}[!htbp]
\centering
\includegraphics[width=\textwidth]{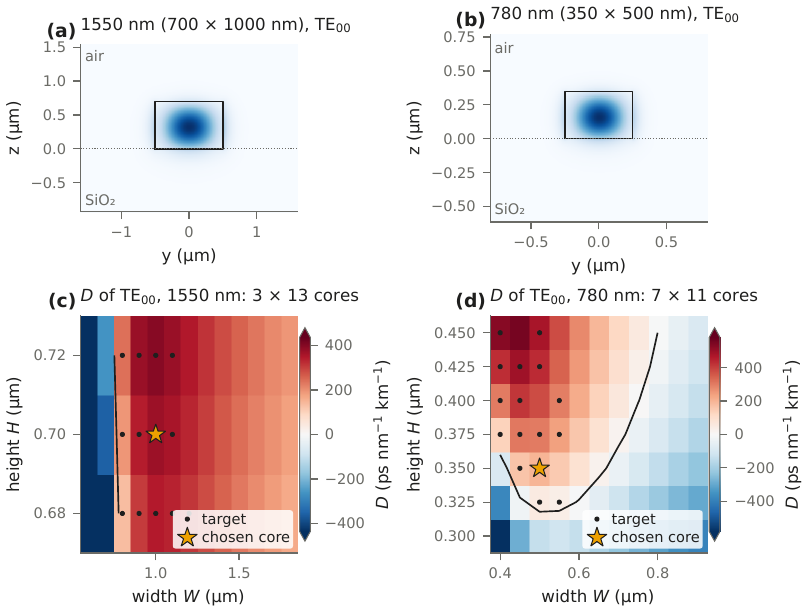}
\caption{(a, b) \TE{} mode of the chosen cores, $700\times1000$\,nm at 1550\,nm and $350\times500$\,nm at 780\,nm, air-clad on \SiO. (c, d) $D$ of \TE{} over the map of core heights and widths (50 steps per wavelength, two-sided average). Dots: targets (anomalous, single TE mode). Star: chosen core. Black line: zero dispersion.}\label{fig:xs}
\end{figure}

\paragraph{Dispersion and target rule} The core, oxide and silicon handle indices are tabulated bulk data~\cite{luke2015,malitson1965,franta2017}, fitted with a pole-residue model over the design band (Supplementary Section~S1). As in the finite-difference eigenmode analysis that sizes single-mode \SiN{} waveguides and microrings~\cite{mandal2023valley,mandal2024emission,mandal2026stitchfree}, for each cross-section the solver finds the guided modes at nine wavelengths over $\pm0.1$\um{} around the pump, and a fourth-order polynomial fit of $\neff(\lambda)$ gives
\begin{equation}
n_g = \neff - \lambda \frac{\mathrm{d}\neff}{\mathrm{d}\lambda}, \qquad D = -\frac{\lambda}{c} \frac{\mathrm{d}^2 \neff}{\mathrm{d}\lambda^2}.
\end{equation}
A cross-section is a \emph{target} when its pump mode (\TE{} by default) has $D > 0$, is the only guided mode of its polarisation, and keeps both over the fabrication tolerance box of $\pm50$\,nm in width and $\pm20$\,nm (1550\,nm) or $\pm10$\,nm (780\,nm) in film thickness. TARANG maps $D$ over heights and widths (Fig.~\ref{fig:xs}c, d) and ranks the targets by how many of their eight neighbours on the map are also targets, then by the \TE--\TM{} index split, which keeps the two families' resonances apart.

\begin{figure}[!htbp]
\centering
\includegraphics[width=\textwidth]{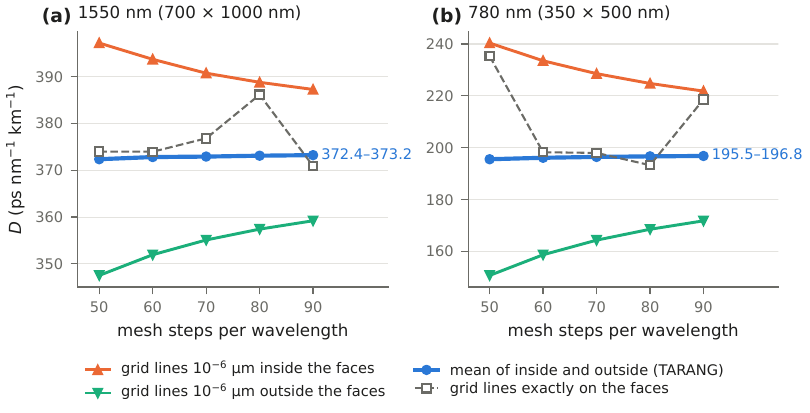}
\caption{$D$ against the mesh for the two chosen cores. With the grid lines exactly on the core's faces (grey), field samples lie on the faces and their material is set by floating-point rounding, so $D$ jumps between meshes. With the lines $10^{-6}$\um{} inside (orange) or outside (green) the faces the choice is definite; the two converge from opposite sides with an error of first order in the cell size, and their mean (blue), used by TARANG, changes by less than 1\Dunit{} at 1550\,nm and 1.3\Dunit{} at 780\,nm over the ladder.}\label{fig:snap}
\end{figure}

\paragraph{Face sampling and the two-sided average} Because $D$ is a second derivative, a change of $10^{-5}$ in $\neff$ at one wavelength moves it by tens of\Dunit. A common way to steady it is to put grid lines exactly through the core's faces, so that the faces fall in the same place in their cells on every mesh. TARANG runs Tidy3D's finite-difference mode solver locally, where by default it does not apply subpixel averaging (the remote solver does, as does the optional \texttt{tidy3d-extras} package), so each field sample gets one material. With a grid line on a face, the samples of the field components parallel to that face lie \emph{on} it (on the Yee grid the normal component sits half a cell away), and whether they see the core or the cladding is decided by floating point rounding. On the 780\,nm core a $10^{-12}$\um{} change of the width moved $\neff$ by $3\times10^{-3}$, and the same calculation gave $D = 198.8$ and 235.3\Dunit{} on two computers. TARANG therefore solves each cross-section twice, with the grid lines $10^{-6}$\um{} inside and $10^{-6}$\um{} outside every face, and averages the two. 
These two computations yield errors of opposite sign but nearly identical magnitude, thereby cancelling out the first-order error (Fig.~\ref{fig:snap}). Over 50--90 mesh steps per wavelength the mean stays within 372.4--373.2\Dunit{} at 1550\,nm and 195.5--196.8\Dunit{} at 780\,nm, while the solve with the lines on the faces scatters by up to 42\Dunit. TARANG quotes $D$ as the median over this five-mesh ladder, with the largest distance from it as its band. Subpixel averaging of the permittivity~\cite{farjadpour2006}, applied by Tidy3D's cloud mode solver, removes the effect at its root: on a snapped grid it gives the samples on the faces the mean of the two permittivities, and since $\neff$ responds linearly to a small change of permittivity, the two-sided mean is its first-order equivalent. Solved in the cloud with subpixel averaging, both cores give $D$ within 0.35\Dunit{} of the two-sided mean at every mesh of the ladder (Supplementary Section~S2). TARANG keeps the two-sided mean so that maps of hundreds of cross-sections run free on the local solver.

\paragraph{Results} On a 700\,nm film the most robust target is $700\times1000$\,nm (Fig.~\ref{fig:xs}a, c), with $D = 372.9 \pm 0.6$\Dunit, $n_g = 2.2227$, and $D$ between 335.7 and 397.0\Dunit{} over the tolerance box, which is anomalous and single-mode throughout. On a 350\,nm film the chosen target is $350\times500$\,nm (Fig.~\ref{fig:xs}b, d), with $D = 196.4 \pm 0.9$\Dunit, $n_g = 2.2692$, and $D$ between 97.1 and 241.9\Dunit{} over the tolerance box. On both platforms the mesh band is far smaller than the fabrication spread, so the tolerance, not the numerics, limits how well $D$ is known before a chip is measured.

%% file: sections/04_ring.tex
\section{Ring and coupling}\label{sec:ring}

\begin{figure}[!htbp]
\centering
\includegraphics[width=\textwidth]{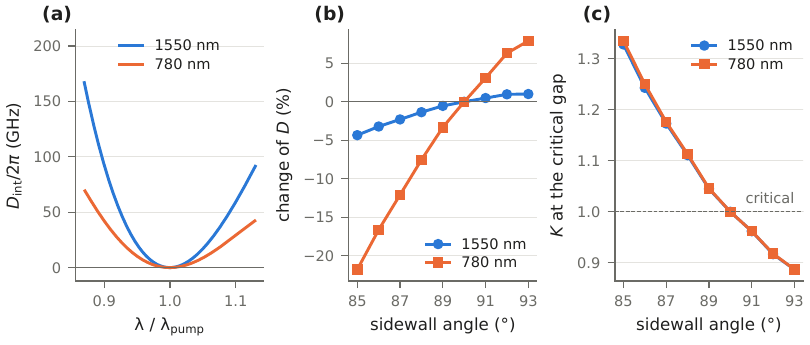}
\caption{(a) Integrated dispersion of the two 200\,GHz rings ($D_2/2\pi = 16.07$ and 2.10\,MHz). (b) Change of $D$ and (c) the coupling ratio $K$ at the designed critical gap against the etched sidewall angle, with the widths fixed at the top of the core.}\label{fig:ring}
\end{figure}

\paragraph{Radius and $Q$} The ring radius follows from the FSR as $R = c / (2\pi n_g\,\mathrm{FSR})$, giving 107.5\um{} for 200\,GHz at 1550\,nm (199.7\,GHz). The losses are assumed values with published sources. Their total, 0.4\,dB/cm, matches the intrinsic $Q$ of about $10^6$ measured on electron-beam-patterned, annealed LPCVD \SiN{} rings~\cite{colacion2026}; a local solve of the bent mode adds a bend loss below 0.001\,dB/cm. The intrinsic $Q$ follows from the power attenuation coefficient $\alpha$~\cite{bogaerts2012},
\begin{equation}
Q_i = \frac{2 \pi n_g}{\lambda \alpha},
\end{equation}
and is $9.8\times10^5$ at 1550\,nm. A broadband solve of the bent waveguide gives the integrated dispersion $\Dint$; at 1550\,nm $D_2/2\pi = 16.07$\,MHz (Fig.~\ref{fig:ring}a).

\begin{figure}[!htbp]
\centering
\includegraphics[width=\textwidth]{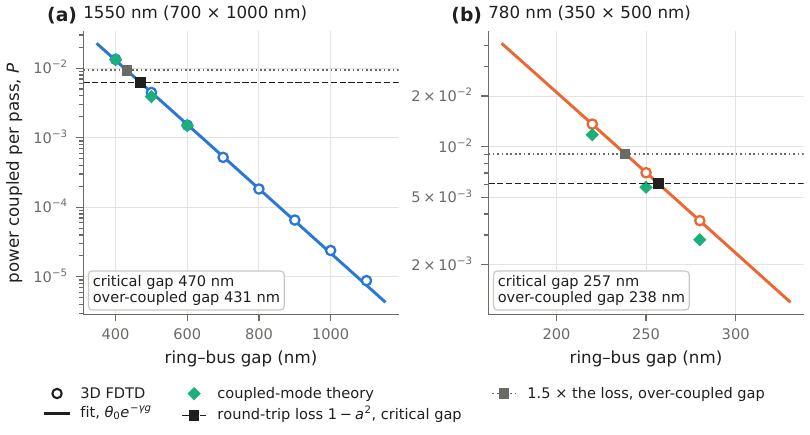}
\caption{Power coupled into the ring per pass, $P = \sin^2\theta$, against the ring--bus gap, from 3D FDTD of a 45$^\circ$ arc beside the bus (circles), the exponential fit of $\theta$ (line) and the coupled-mode estimate from local supermodes (diamonds). Dashed: the round-trip power loss $1 - a^2$; the ring is critically coupled at the gap where the coupled power equals it (black square). Dotted: 1.5 times the loss, the over-coupled design (grey square).}\label{fig:gap}
\end{figure}

\paragraph{Ring--bus gap} TARANG simulates a 45$^\circ$ arc of the ring beside the straight bus in 3D FDTD at several gaps $g$, records the power $P$ that enters the arc, and fits the coupling angle per pass, $\theta = \arcsin\sqrt{P}$, with $\theta = \theta_0 \exp(-\gamma g)$. The critical gap $g_c$ is where the coupling equals the round-trip loss~\cite{bogaerts2012},
\begin{equation}
\sin^2 \theta(g_c) = 1 - a^2 ,
\end{equation}
with $a^2$ the round-trip power survival set by $Q_i$. Their ratio, the coupling ratio $K = \sin^2\theta/(1-a^2) \approx Q_i/Q_c$, is 1 at critical coupling, above 1 for an over-coupled and below 1 for an under-coupled ring, and sets the transmission at resonance, $T_0 = [(1-K)/(1+K)]^2$. An over-coupled design takes $K = 1.5$, for a dip less sensitive to fabrication. As an independent check at no cloud cost, the even and odd local supermodes of two cores side by side give the coupling per length $\kappa = \pi\Delta n/\lambda$~\cite{yariv1973}, and integrating it along the ring, where the gap grows as $x^2/2R$, gives
\begin{equation}
\theta = \kappa(g) \sqrt{\frac{2 \pi R}{\gamma}} .
\end{equation}
At 1550\,nm the FDTD fit ($R^2$ of $\ln\theta$ = 0.99985) gives $g_c = 470$\,nm and $\gamma = 5.32$\,\textmu m$^{-1}$, and coupled-mode theory 465\,nm and 5.47\,\textmu m$^{-1}$; at 780\,nm 257 against 248\,nm (Fig.~\ref{fig:gap}). At the critical gap 0.62\pct{} of the power crosses into the ring per pass; the vanishing transmission of critical coupling builds up only at resonance over many round trips, and is computed from this per pass coupling in the whole device model (Supplementary Section~S6). The FDTD field maps are in Supplementary Fig.~S1.

\paragraph{Tolerance} The width tolerance moves the gap one for one. At the critical 1550\,nm design, waveguides 50\,nm wider or narrower give $K = 1.70$ or 0.59, a resonance dip of 11.7\,dB instead of more than 40\,dB (Supplementary Table~S3), so the mask carries ring variants around both the critical and the over-coupled gap. A \TM{} family crosses \TE{} about every 880 modes, so 0.3 crossings fall in the 1450--1650\,nm band.

\paragraph{Sidewall angle} TARANG keeps the widths at the top of the core and solves sloped walls from 85$^\circ$ to 93$^\circ$. $D$ changes by $-4.3$\pct{} (85$^\circ$) to $+1.0$\pct{} (93$^\circ$) at 1550\,nm and by $-21.8$\pct{} to $+7.9$\pct{} at 780\,nm (Fig.~\ref{fig:ring}b). At the critical gap, $K$ goes from 1.33 at 85$^\circ$ to 0.89 at 93$^\circ$, the same on both platforms to within 0.01 (Fig.~\ref{fig:ring}c). This is a scaling. A sloped wall narrows the gap at the bottom of the core by an amount set by $H/\tan\alpha$, which scales with the film thickness $H$ and so with $\lambda$, while the coupling decays as $\exp(-\gamma g)$ with $\gamma \propto 1/\lambda$; the product $\gamma H/\tan\alpha$, which sets the change of $K$, is the same on both platforms.

%% file: sections/05_gratings.tex
\section{Grating couplers}\label{sec:grating}

\paragraph{Method} Light enters and leaves the chip through fully etched grating couplers of 30 teeth, simulated in 2D FDTD with a Gaussian fibre beam at 12$^\circ$ from vertical. The first-order phase-matching condition~\cite{taillaert2006}, $\Lambda = \lambda / (n_\mathrm{avg} - \sin\theta)$, gives a starting period for each fill factor and TARANG maps period $\times$ fill factor around it and then tunes the period onto the design wavelength, refines the fill factor and the fibre position, and tries to centre the efficiency peak, keeping the centred design only if it costs less than 0.5\,dB. The grating width follows from the lateral overlap of the fibre and slab modes, and the taper length from a 3D EME run. Two variants are built on it, an apodised grating whose teeth follow the scattering strength of uniform gratings so that the outgoing beam takes the fibre's Gaussian shape, and a focusing grating with curved teeth that replaces the long taper.

\begin{figure}[!htbp]
\centering
\includegraphics[width=\textwidth]{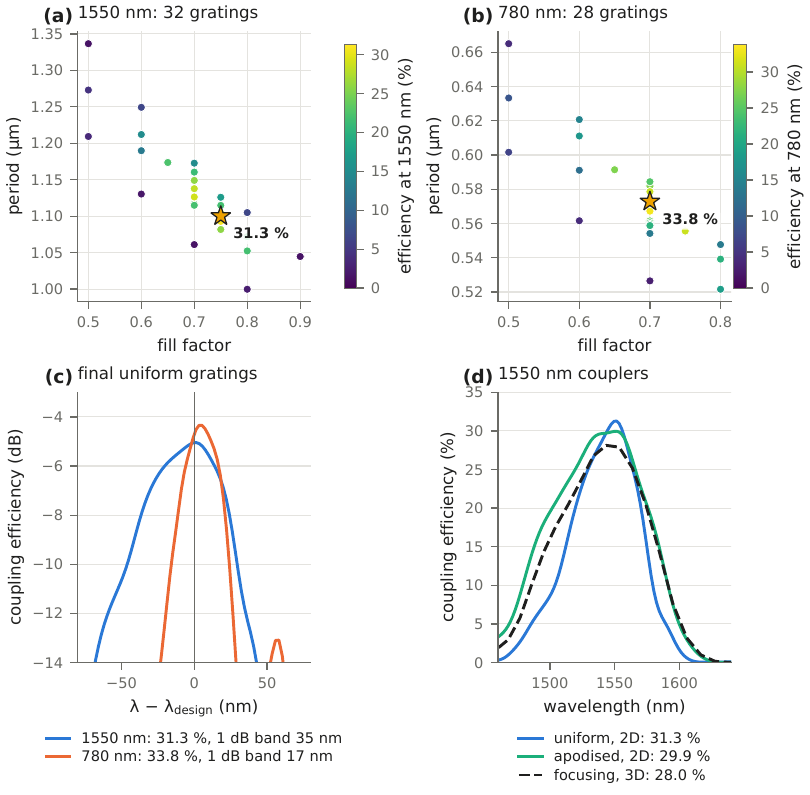}
\caption{Grating couplers (full etch, 30 teeth, fibre at 12$^\circ$). (a, b) All gratings simulated during the design, coloured by the efficiency at the design wavelength; star: final design. (c) Final uniform gratings against the detuning from their design wavelengths (1550 and 780\,nm), so that both share one axis: 1\,dB bandwidths of 35 and 17\,nm, the same 2.2--2.3\pct{} of the wavelength; the 780\,nm peak lies 4.3\,nm above the design wavelength. (d) Uniform and apodised gratings at 1550\,nm (2D) and the focusing grating with apodised teeth (3D FDTD). All at the design mesh; the converged 1550\,nm value is 27.4\pct{} (Supplementary Fig.~S2).}\label{fig:gc}
\end{figure}

\paragraph{1550\,nm} The uniform grating (period 1.1001\um, fill factor 0.75; Fig.~\ref{fig:gc}a, c) couples 31.3\pct{} at the design mesh, with a 1\,dB bandwidth of 35\,nm; it is 14.5\um{} wide and needs a 290\um{} taper (0.048\,dB). A mesh ladder from 15 to 45 cells per wavelength converges monotonically to 27.4\pct{} ($-5.62$\,dB), the two finest meshes differing by 0.04\,dB, so the design mesh overestimates the efficiency by 0.57\,dB but places the peak within 5\,nm (Supplementary Section~S4). The apodised grating, with every trench wider than 200\,nm, reaches 29.9\pct{} with a 1\,dB bandwidth of 53\,nm (Fig.~\ref{fig:gc}d). With these teeth the focusing grating shortens the coupler from 310 to 33\um; a 3D FDTD run gives 28.0\pct{} with a 50\,nm 1\,dB bandwidth.

\paragraph{780\,nm} This grating is harder to simulate than the 1550\,nm one. At 15 cells per wavelength a 1.5 times finer mesh still moved its efficiency by 1.2\,dB, and on finer meshes the simulations with a perfectly matched layer (PML) diverged, because dispersive materials, including the absorbing silicon handle, reach the boundary. It is therefore designed at 30 cells per wavelength with the stable PML. A second boundary, an adiabatic absorber, stays stable on finer meshes but adds reflection lobes about 15\,nm apart; averaged over one lobe period, the two boundaries agree to 0.2\,dB and the two finest meshes to 0.01\,dB (Supplementary Section~S4). The uniform grating (period 0.5729\um, fill factor 0.70; Fig.~\ref{fig:gc}b, c) couples 33.8\pct{} ($-4.72$\,dB) at 780\,nm, with a peak of 36.8\pct{} at 784.3\,nm and a 1\,dB bandwidth of 17\,nm; shortening the period to move the peak towards 780\,nm lowered the efficiency there by 1.4\,dB or more, while tilting the fibre to 14$^\circ$ centres it at 34.0\pct. The focusing grating shortens the coupler from 157 to 30\um. No apodised grating could be designed under a full etch. The teeth that could be simulated (fill factors 0.55--0.65) scatter over 1.6--1.9\um, less than the 2.5\um{} fibre waist, and the weaker teeth that apodisation needs (fill factors 0.70--0.79, trenches down to the 120\,nm the process can clear) did not give stable simulations (Supplementary Section~S4).

\paragraph{Tolerance} The sidewall angle matters most. A tooth wall 3$^\circ$ off vertical costs more than 2\,dB on both platforms, against about 0.5\,dB or less for $\pm1^\circ$ of fibre angle or the film-thickness tolerance (Supplementary Section~S5). The mask therefore carries grating variants with the teeth $\pm10$ and $\pm20$\,nm and the period $\pm20$\,nm (1550\,nm) or $\pm10$\,nm (780\,nm).

%% file: sections/06_comparison.tex
\section{The two platforms compared}\label{sec:compare}

Table~\ref{tab:compare} sets the two designs side by side. Three differences follow from halving the wavelength at the same FSR. First, the same loss in dB/cm gives twice the intrinsic $Q$, and with the smaller mode area the pump needed for four times threshold drops from 208 to 25\,mW. Second, $D_2$ is almost eight times smaller, which gives a broader comb and a shorter soliton. Third, the evanescent field decays twice as fast across the gap, so the coupling is twice as sensitive to fabrication. The width tolerance that moves $K$ between 0.59 and 1.70 at 1550\,nm moves it between 0.33 and 3.00 at 780\,nm, and a gap and dose matrix on the mask is essential there. The thinner film also makes $D$ less certain; over the tolerance box it spans a factor of 2.5 at 780\,nm against 1.2 at 1550\,nm, mostly through the film thickness ($\mathrm{d}D/\mathrm{d}H \approx 4.9$\Dunit{} per nm at 780\,nm).

\begin{table}[!htbp]
\centering
\caption{The two designs side by side. Grating efficiencies are 2D FDTD at the design mesh unless stated; the whole-device model is in Supplementary Section~S6, the pyLLE simulations in Section~S7.}\label{tab:compare}
\small
\begin{tabularx}{\textwidth}{>{\raggedright\arraybackslash}p{4.6cm}LL}
\toprule
Quantity & 1550\,nm & 780\,nm \\
\midrule
Core ($H\times W$) & $700\times1000$\,nm & $350\times500$\,nm \\
$D$ (five-mesh ladder) & $372.9 \pm 0.6$\Dunit & $196.4 \pm 0.9$\Dunit \\
$D$ over the tolerance box & 335.7 to 397.0\Dunit & 97.1 to 241.9\Dunit \\
$n_g$ & 2.2227 & 2.2692 \\
Ring radius, FSR & 107.5\um, 199.7\,GHz & 105.0\um, 200.3\,GHz \\
$D_2/2\pi$ & 16.07\,MHz & 2.10\,MHz \\
$Q_i$ at 0.4\,dB/cm & $9.8\times10^5$ & $2.0\times10^6$ \\
Critical / over-coupled gap & 470 / 431\,nm & 257 / 238\,nm \\
Gap decay rate $\gamma$ (3D FDTD) & 5.32\,\textmu m$^{-1}$ & 11.01\,\textmu m$^{-1}$ \\
$K$ at $\pm50$\,nm of width (critical design) & 1.70 / 0.59 & 3.00 / 0.33 \\
Uniform grating & 31.3\pct; 27.4\pct{} at the converged mesh & 33.8\pct \\
Focusing grating & 28.0\pct{} (3D FDTD) & 33.2\pct{} (estimate) \\
Fibre to fibre, off resonance & $-11.1$\,dB & $-9.6$\,dB \\
Pump for 4$\times$ threshold & 208\,mW & 25\,mW \\
Single soliton: pulse, comb power & 130\,fs, 16.9\,mW & 47\,fs, 0.78\,mW \\
\bottomrule
\end{tabularx}
\end{table}

%% file: sections/07_validation.tex
\section{Validation}\label{sec:validation}

\begin{figure}[!htbp]
\centering
\includegraphics[width=\textwidth]{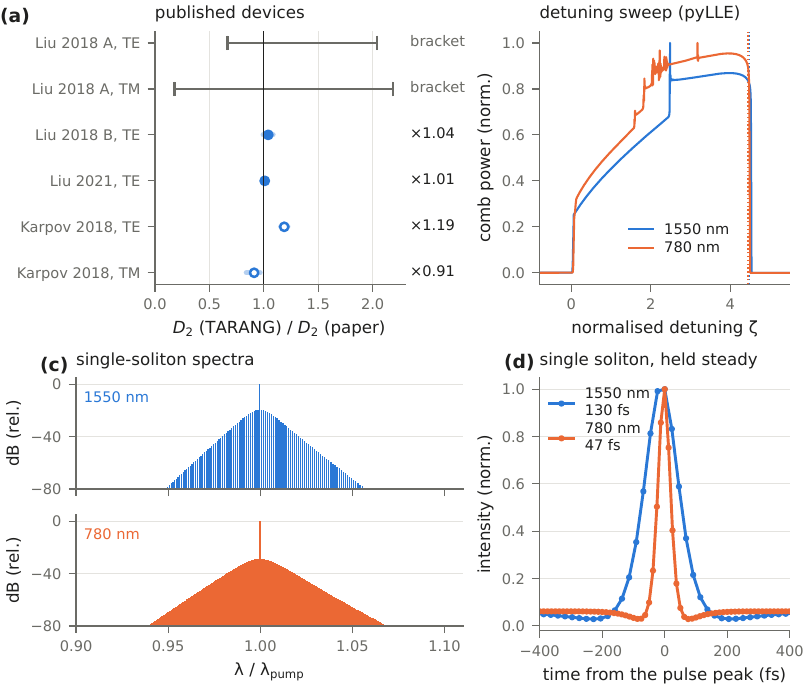}
\caption{Validation. (a) Ratio of TARANG's $D_2$ to the published value for the six comparisons that report a value (the seventh, Moille et al., reports only that $D$ is anomalous at every width, and is in Table~\ref{tab:lit}); filled: measured, open: simulated; bars: film $\pm10$\,nm; Damascene device: bracket between all-\SiO{} and all-air top cladding. (b) pyLLE detuning sweep from TARANG's exported parameters at four times threshold (pump 208\,mW at 1550\,nm and 25\,mW at 780\,nm), with soliton steps up to the existence limit $\zeta_\mathrm{max}$ (dotted). (c) Spectra and (d) intensity of a single soliton held at $\zeta = 3.11$ (1550\,nm) and 3.09 (780\,nm).}\label{fig:validation}
\end{figure}

\paragraph{Published devices} TARANG solved the cross-sections of five published \SiN{} microresonators from four papers~\cite{liu2018,liu2021,karpov2018,moille2025}, two of them in both \TE{} and \TM{}, seven comparisons in all (Table~\ref{tab:lit}, Fig.~\ref{fig:validation}a). They cover pumps from 780 to 1596\,nm, air and \SiO{} cladding, a Damascene device and a 23\um{} ring; TARANG used the papers' own dimensions and computed $D_2 = -(c/n_g)\,D_1^2\,\beta_2$ from the papers' FSRs~\cite{herr2014}. The three devices with a single published value (four comparisons) were also solved with the film 10\,nm thinner and thicker, the usual run-to-run spread of LPCVD films. For the two measured devices with a plain stack, TARANG is within 1\pct{} and 4\pct{} of the measurement, inside the film-thickness spread. The Damascene device (oxide beside the core, air above) cannot be modelled exactly with a uniform top cladding, and its measured values lie between TARANG's all-air and all-\SiO{} limits. For the air-clad rings of Moille et al., which stay anomalous at every width from 575 to 775\,nm on a 430\,nm film, TARANG gives $D$ between 147 and 167\Dunit{} (\TM), anomalous at every width. The largest differences are with Karpov et al., whose values are themselves simulations of a ring with 80$^\circ$ sidewalls and an unstated cladding, against vertical walls and \SiO{} here. Solved with 80$^\circ$ walls, the same core gives $D_2/2\pi$ between $-57$ and $-18$\,MHz (\TE) and between $-1.7$ and 24\,MHz (\TM), depending on whether the stated width is that of the top, the middle or the bottom of the core, which the paper does not specify; both published values lie inside these ranges (Supplementary Section~S8).

\begin{table}[!htbp]
\centering
\caption{$D_2/2\pi$ from TARANG against published \SiN{} microresonators. TARANG uses vertical walls; for the Karpov et al. ring with its 80$^\circ$ walls see Supplementary Section~S8. Damascene device: the measured value against TARANG's all-\SiO{} and all-air limits; Moille et al.: the sign of $D$ across the stated widths; ---: not applicable to these two kinds of test.}\label{tab:lit}
\footnotesize
\setlength{\tabcolsep}{3.5pt}
\begin{tabularx}{\textwidth}{>{\raggedright\arraybackslash}p{3.3cm}l>{\raggedright\arraybackslash}p{1.65cm}L>{\raggedright\arraybackslash}p{1.6cm}>{\raggedright\arraybackslash}p{1.9cm}}
\toprule
Device & Mode & Paper & TARANG & Difference & TARANG, film $\pm10$\,nm \\
\midrule
Liu et al.~\cite{liu2018}, device B (\SiO{} clad), measured & \TE & 1.23\,MHz & 1.28\,MHz & $+4.1$\pct & 1.22--1.33\,MHz (inside) \\
Liu et al.~\cite{liu2021} (\SiO{} clad), measured & \TE & 0.224\,MHz & 0.226\,MHz & $+0.9$\pct & 0.222--0.229\,MHz (inside) \\
Liu et al.~\cite{liu2018}, device A (Damascene), measured & \TE & 1.10\,MHz & 0.74\,MHz (all \SiO) to 2.25\,MHz (all air) & bracketed & --- \\
Liu et al.~\cite{liu2018}, device A (Damascene), measured & \TM & 0.967\,MHz & 0.17\,MHz (all \SiO) to 2.12\,MHz (all air) & bracketed & --- \\
Karpov et al.~\cite{karpov2018}, $R\approx23$\um, simulated & \TE & $-32$\,MHz & $-38.0$\,MHz & $-18.8$\pct & $-39.1$ to $-37.0$\,MHz \\
Karpov et al.~\cite{karpov2018}, $R\approx23$\um, simulated & \TM & 11\,MHz & 10.0\,MHz & $-8.9$\pct & 9.2--10.6\,MHz \\
Moille et al.~\cite{moille2025}, widths 575--775\,nm on 430\,nm & \TM & anomalous at every width & $D = 147$--167\Dunit, anomalous at every width & agrees & --- \\
\bottomrule
\end{tabularx}
\end{table}

\paragraph{Soliton formation} The integrated dispersion and ring parameters exported by TARANG were handed to pyLLE~\cite{moille2019} and pumped at four times threshold (normalised pump $f^2 = 3.64$ and 3.60). On both designs a detuning sweep shows the soliton steps up to the existence limit $\zeta_\mathrm{max} = \pi^2 f^2/8$~\cite{herr2014}, and a single soliton seeded at $\zeta = 3.11$ (1550\,nm) and 3.09 (780\,nm) stays one steady pulse over $2\times10^4$ round trips (Fig.~\ref{fig:validation}b--d), with a full width at half maximum of 130 and 47\,fs (analytic estimates 144 and 53\,fs). Every simulated mode lies inside the computed range of $\Dint$, so pyLLE extrapolates nothing (Supplementary Section~S7).

\paragraph{Numerical checks} Each quoted number has its own check. $D$ is the median over a five-mesh ladder with a band under 1\Dunit, and the two-sided solve gives the same $\neff$ on Linux and Windows to nine decimals; the 1550\,nm grating efficiency is quoted at the converged mesh, and the 780\,nm grating is checked against a second boundary; and the ring--bus gap from FDTD agrees with coupled-mode theory to 5\,nm at 1550\,nm and 9\,nm at 780\,nm.

%% file: sections/08_conclusions.tex
\section{Conclusions}\label{sec:conclusion}

TARANG turns the design of an air-clad, fully etched silicon nitride microcomb chip into a single recorded run, from tabulated refractive indices to a mask, with every cloud simulation priced and capped, every result cached by its content, and every quoted number checked numerically. Two complete designs show what moving from 1550 to 780\,nm changes. The threshold falls eightfold and the soliton shortens, but the thinner film makes the dispersion and the coupling far more sensitive to fabrication.

Three lessons apply beyond this package. First, snapping the grid onto a core's faces steadies a dispersion calculation but leaves the samples on the faces to rounding; averaging the solves just inside and just outside the faces gives the same $D$ on every computer we tested, steady to about 1\Dunit{} from mesh to mesh. Second, an FDTD result with dispersive materials reaching the absorbing boundary must be checked against a second boundary and on several meshes; the design mesh overestimated the efficiency of the 1550\,nm grating by 0.6\,dB. Third, the etch depth sets which gratings can be made. With a full etch the 1550\,nm design carries uniform, apodised and focusing gratings; at 780\,nm every full-etch tooth scatters strongly, over 1.6--1.9\um{}, less than the 2.5\um{} fibre waist, so the design keeps the uniform and focusing gratings, and a partial etch, already a setting in TARANG, would give the weaker teeth that apodisation needs.

Both designs are ready for fabrication with one electron-beam lithography step and one full etch. The layout carries ring variants at $\pm50$\,nm (1550\,nm) and $\pm25$\,nm (780\,nm) of gap around the critical and over-coupled designs, grating variants in tooth length and period, and a reference waveguide. Measured resonance frequencies, quality factors, coupling and grating spectra of the first chips will test the predictions directly, and the measured losses can replace the assumed ones in the design.

%% file: sections/09_back_matter.tex
\section*{Acknowledgements}

The simulations used Tidy3D (Flexcompute Inc.)~\cite{tidy3d}: its finite-difference eigenmode solver, run locally, for every cross-section and bent-waveguide mode solve; its cloud FDTD solver, with perfectly matched layer, stable PML and adiabatic absorber boundaries, for the ring--bus couplers and the grating couplers; its eigenmode-expansion solver for the tapers; and its \texttt{FastDispersionFitter} for the pole-residue fits of the materials. The soliton simulations used pyLLE~\cite{moille2019}.

This work was supported by the Department of Science and Technology (DST), Government of India, under the National Quantum Mission through the Quantum Materials and Devices (QMD) Hub.

\section*{Code availability}

TARANG is released under the MIT licence. All results in this paper were computed with TARANG v0.1.5~\cite{tarang_zenodo} and Tidy3D 2.12.0~\cite{tidy3d}. Parts of TARANG's code, tests and documentation were written with an AI coding assistant (Claude, Anthropic) under the authors' direction, and each such commit records this; the authors reviewed and tested all code, ran every simulation reported here, and take full responsibility for the software and the results. About 590 tests cover the physics formulas, the geometry, the cache, the cost guard and each stage, and run on Linux and Windows.

\section*{Data availability}

The design folders of both designs, with every setting, result and figure data file, and the scripts that draw every figure, will be deposited on Zenodo with a DOI of their own, linked from the software record~\cite{tarang_zenodo}.

\section*{Declaration of generative AI and AI-assisted technologies in the manuscript preparation process}

During the preparation of this work the authors used Claude (Anthropic) in order to edit the text of the manuscript and to write the scripts that draw the figures from the saved design results. After using this tool, the authors reviewed and edited the content as needed and take full responsibility for the content of the published article.

\section*{Declaration of competing interest}

The authors declare that they have no known competing financial interests or personal relationships that could have appeared to influence the work reported in this paper.

%% file: sections_supplementary/S1_materials.tex
\section{Materials}\label{sec:s-materials}

The core, buried oxide and handle are described by tabulated bulk refractive-index data shipped with TARANG: stoichiometric \SiN{} from Luke et al.~\cite{luke2015}, fused silica from Malitson~\cite{malitson1965}, and float-zone silicon at 300\,K from Franta et al.~\cite{franta2017}. Each material is fitted with a pole-residue model over the design band, which the FDTD solver needs. For pumps below 1.1\um{} the fit covers the band plus 5\pct{} on each side, so that it is not an extrapolation of a telecom-band fit. The air cladding is non-dispersive. Table~\ref{tab:s-materials} lists the band and the indices at the pump.

These are bulk values. The deposited film's index differs from run to run, which makes the materials the dominant systematic error of every prediction; the design report states this, and a measured film index can replace the tabulated one.

\begin{table}[htbp]
\centering
\caption{Design band and refractive indices at the pump.}\label{tab:s-materials}
\begin{tabular}{lll}
\toprule
 & 1550\,nm design & 780\,nm design \\
\midrule
Band of the fits & 1.348--1.752\um & 0.679--0.881\um \\
\SiN{} core (data 0.31--5.50\um) & 1.9963 & 2.0256 \\
\SiO{} buried oxide (data 0.40--3.00\um) & 1.4440 & 1.4537 \\
Si handle & 3.4795 & 3.6927 \\
\bottomrule
\end{tabular}
\end{table}

%% file: sections_supplementary/S2_cross_section.tex
\section{Cross-section numerics}\label{sec:s-xs}

\paragraph{Mesh ladder} Table~\ref{tab:s-ladder} gives the full ladders behind Fig.~3. TARANG quotes the median of the two-sided mean over the five meshes (the 70-step value) and the largest distance from it as the band, giving $372.9 \pm 0.6$ and $196.4 \pm 0.9$\Dunit. The inside and outside solves differ by 50 and 90\Dunit{} at 50 steps and close in as the mesh is refined, as a first-order error in the cell size should; their mean changes by less than 1.3\Dunit{} over the ladder. With the lines on the faces the result depends on how each face coordinate rounds on a given mesh; at 780\,nm it reads 235.3, 198.2, 198.0, 193.2 and 218.6\Dunit.

\begin{table}[htbp]
\centering
\caption{$D$ of \TE{} (\Dunit) against the mesh (steps per wavelength in the core). Inside, outside: grid lines $10^{-6}$\um{} inside or outside every face; mean: their average, used by TARANG; on faces: grid lines exactly on the faces, computed on the design computer (Windows); the local solver gives each sample one material. Subpixel: Tidy3D's cloud mode solver with subpixel averaging of the permittivity, with the grid lines on the faces (snapped) or without snapping (free).}\label{tab:s-ladder}
\small
\begin{tabular}{rrrrr|rr}
\toprule
 & \multicolumn{4}{c|}{Local solver} & \multicolumn{2}{c}{Subpixel (cloud)} \\
Steps & inside & outside & mean & on faces & snapped & free \\
\midrule
\multicolumn{7}{l}{1550\,nm, $700\times1000$\,nm} \\
50 & 397.2 & 347.5 & 372.4 & 374.0 & 372.4 & 370.8 \\
60 & 393.7 & 352.0 & 372.8 & 374.0 & 372.7 & 376.9 \\
70 & 390.7 & 355.1 & 372.9 & 376.8 & 372.9 & 370.5 \\
80 & 388.8 & 357.4 & 373.1 & 386.1 & 373.0 & 376.1 \\
90 & 387.3 & 359.2 & 373.2 & 370.9 & 373.0 & 372.0 \\
\midrule
\multicolumn{7}{l}{780\,nm, $350\times500$\,nm} \\
50 & 240.4 & 150.7 & 195.5 & 235.3 & 195.2 & 191.7 \\
60 & 233.5 & 158.7 & 196.1 & 198.2 & 195.7 & 192.8 \\
70 & 228.6 & 164.3 & 196.4 & 198.0 & 196.1 & 193.9 \\
80 & 224.8 & 168.5 & 196.6 & 193.2 & 196.3 & 194.9 \\
90 & 221.8 & 171.8 & 196.8 & 218.6 & 196.5 & 195.8 \\
\bottomrule
\end{tabular}
\end{table}

\paragraph{Subpixel averaging} Tidy3D's cloud mode solver applies subpixel averaging of the permittivity~\cite{farjadpour2006}. Solved there on the same five meshes with the grid lines on the faces, the two cores give $D$ within 0.2\Dunit{} (1550\,nm) and 0.35\Dunit{} (780\,nm) of TARANG's two-sided mean at every mesh, with the same slow drift over the ladder (Table~\ref{tab:s-ladder}). The two remedies therefore agree; the residual 0.3\Dunit{} at 780\,nm is a third of the mesh band. Without snapping, subpixel averaging alone still varies by 6.4 and 4.1\Dunit{} over the ladder, because the faces then fall at a different place in their cells on each mesh. The 20 cloud solves cost 1.4 FlexCredits; the script and its results are archived with the design folders.

\paragraph{Computer-to-computer check} With the lines on the faces, the 780\,nm core at 50 steps gave $D = 198.8$\Dunit{} on Linux and 235.3\Dunit{} on Windows, from the same code and grids equal to $10^{-13}$\um. With the two-sided average both computers give $D = 195.53$\Dunit, and the effective indices of the two sides agree to nine decimals. TARANG ships the script that runs this check and the Linux reference values.

\paragraph{Mode properties} The 1550\,nm core has an effective area of 0.746\,\textmu m$^2$, 82\pct{} of the power in the core, and a \TE--\TM{} index split of 0.029; the 780\,nm core 0.183\,\textmu m$^2$, 82\pct{} and 0.031. Each guides one TE and one TM mode. The ring's dispersion profile at 780\,nm crosses zero at 868\,nm, inside the band of the fits; at 1550\,nm it has no zero in the band.

%% file: sections_supplementary/S3_ring_bus_coupling.tex
\section{Ring--bus coupling}\label{sec:s-gap}

\begin{figure}[htbp]
\centering
\includegraphics[width=\textwidth]{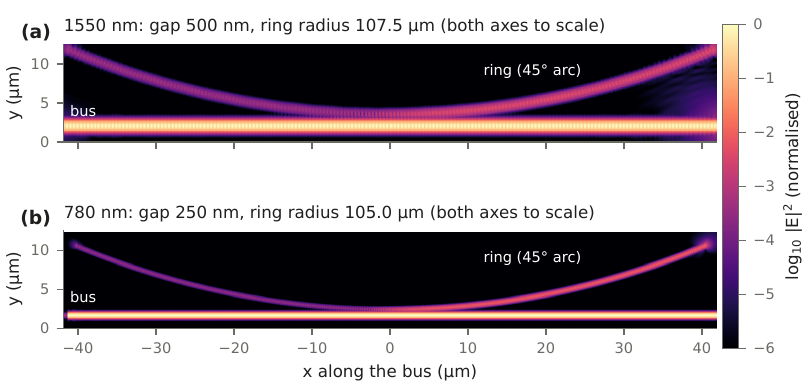}
\caption{$|E|^2$ at the pump wavelength (log scale, normalised to its maximum) in the middle of the core, from the 3D FDTD runs of the ring--bus coupler at (a) 1550\,nm and (b) 780\,nm, at the swept gaps nearest the critical gaps (500 and 250\,nm): one pass of light along the straight bus past a 45$^\circ$ arc of the ring. Both axes are to scale: the arc of a ring of radius about 107\,\textmu m rises only about 8\,\textmu m over its 82\,\textmu m length, and near its lowest point it is the parabola $y \approx x^2/2R$ used in the coupled-mode estimate. The simulation has no closed ring and therefore no resonance; it measures the power that crosses the gap in one pass (0.45\pct{} at 500\,nm and 0.70\pct{} at 250\,nm; the critical gaps, 470 and 257\,nm, lie between swept gaps and come from the fit through all of them, where 0.62\pct{} and 0.61\pct{} cross), so the bus stays bright and the arc is visible only on the logarithmic scale. The vanishing bus transmission of critical coupling is a steady-state effect of the closed ring at resonance and appears in the whole-device spectra (Section~\ref{sec:s-device}).}\label{fig:s-field}
\end{figure}

The FDTD runs cover a 45$^\circ$ arc of the ring beside a straight bus, at gaps of 0.4--1.1\um{} (1550\,nm) and 0.22--0.28\um{} (780\,nm). The power $P$ entering the arc gives the coupling angle per pass, $\theta = \arcsin\sqrt{P}$ (main text, Section 4), and a fit $\theta = \theta_0 \exp(-\gamma g)$ gives the rate $\gamma$ at which the coupling decays with the gap $g$, set by the evanescent tails of the two modes, and the prefactor $\theta_0$, the value extrapolated to zero gap: $\theta_0 = 0.961$ and $\gamma = 5.32$\,\textmu m$^{-1}$ at 1550\,nm ($R^2$ of $\ln\theta$ = 0.99985) and $\theta_0 = 1.319$ and $\gamma = 11.01$\,\textmu m$^{-1}$ at 780\,nm (0.99997). Figure~\ref{fig:s-field} shows the field of the runs. They are single-pass by design. A 3D FDTD run of the whole ring to steady state at $Q_L \approx 5\times10^5$--$10^6$ (photon lifetime about 0.4\,ns on both platforms, about 80 round trips of 5\,ps) would have to cover several hundred round trips on a domain about 40 times larger than the arc's, roughly $10^5$ times the cost of the arc runs. The resonance is instead built from the per-pass coupling in the whole-device model (Section~\ref{sec:s-device}), whose critically coupled rings show dips of 34 and 58\,dB at the pump.

Table~\ref{tab:s-gaptol} gives the effect of the width tolerance. Waveguides 50\,nm wider close the gap by 50\,nm, and the coupling changes by $\exp(\pm\gamma\times 50\,\mathrm{nm})$ in amplitude. At 780\,nm, with twice the decay rate, the critical design moves to $K = 3.0$ or 0.33 and its dip falls to 6\,dB. The over-coupled design is less sensitive in dip depth on the narrow side, which is why both are on the mask.

\begin{table}[htbp]
\centering
\caption{Ring--bus coupling at $\pm50$\,nm of waveguide width ($K$: coupling over round-trip loss; $Q_L$: loaded $Q$).}\label{tab:s-gaptol}
\small
\setlength{\tabcolsep}{4pt}
\begin{tabular}{llrrrr}
\toprule
Design & Case & Gap (nm) & $K$ & $Q_L$ & Dip (dB) \\
\midrule
1550\,nm, critical & 50\,nm wider & 420 & 1.70 & $3.6\times10^5$ & 11.7 \\
 & as designed & 470 & 1.00 & $4.9\times10^5$ & $>40$ \\
 & 50\,nm narrower & 520 & 0.59 & $6.2\times10^5$ & 11.7 \\
1550\,nm, over & 50\,nm wider & 381 & 2.55 & $2.8\times10^5$ & 7.2 \\
 & as designed & 431 & 1.50 & $3.9\times10^5$ & 14.0 \\
 & 50\,nm narrower & 481 & 0.88 & $5.2\times10^5$ & 24.0 \\
780\,nm, critical & 50\,nm wider & 207 & 3.00 & $5.0\times10^5$ & 6.0 \\
 & as designed & 257 & 1.00 & $9.9\times10^5$ & $>40$ \\
 & 50\,nm narrower & 307 & 0.33 & $1.5\times10^6$ & 6.0 \\
780\,nm, over & 50\,nm wider & 188 & 4.49 & $3.6\times10^5$ & 3.9 \\
 & as designed & 238 & 1.50 & $7.9\times10^5$ & 14.0 \\
 & 50\,nm narrower & 288 & 0.50 & $1.3\times10^6$ & 9.5 \\
\bottomrule
\end{tabular}
\end{table}

%% file: sections_supplementary/S4_grating_numerics.tex
\section{Grating numerics}\label{sec:s-gcnum}

\paragraph{1550\,nm mesh ladder} The design runs use 15 cells per wavelength in the core. Table~\ref{tab:s-gcladder} gives the final uniform grating on five meshes. The efficiency falls monotonically and the two finest meshes differ by 0.04\,dB, so the 45-cell value, 27.4\pct, is quoted as converged; the design mesh overestimates it by 0.57\,dB and places the peak 5\,nm too long, well inside the 35\,nm band. The apodised (29.9\pct) and focusing (28.0\pct) results were computed at the design mesh and carry a correction of the same order.

\begin{table}[htbp]
\centering
\caption{The 1550\,nm uniform grating on five meshes (stable PML).}\label{tab:s-gcladder}
\begin{tabular}{rrrrr}
\toprule
Cells per $\lambda$ & Cell (nm) & $\eta$(1550\,nm) & Peak (nm) & Change vs finest (dB) \\
\midrule
15 & 51.7 & 31.3\pct & 1550.7 & 0.57 \\
22 & 35.2 & 29.1\pct & 1547.8 & 0.27 \\
30 & 25.8 & 28.2\pct & 1546.7 & 0.13 \\
38 & 20.4 & 27.7\pct & 1546.0 & 0.04 \\
45 & 17.2 & 27.4\pct & 1545.8 & 0 \\
\bottomrule
\end{tabular}
\end{table}

\begin{figure}[htbp]
\centering
\includegraphics[width=\textwidth]{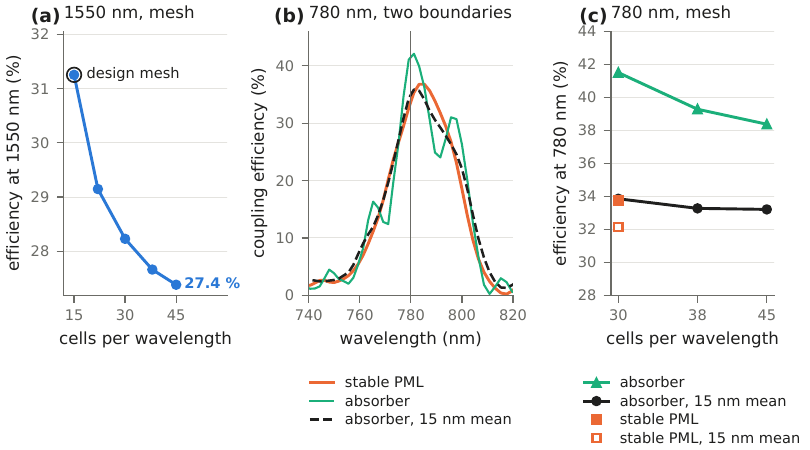}
\caption{Numerical checks of the gratings. (a) 1550\,nm: the final grating on five meshes converges monotonically to 27.4\pct. (b) 780\,nm, same grating and mesh with two absorbing boundaries: the absorber adds lobes about 15\,nm apart that the stable PML does not show; averaged over one lobe period they disappear. (c) Efficiency at 780\,nm against the mesh, raw and averaged over one lobe period.}\label{fig:s-gcnum}
\end{figure}

\paragraph{780\,nm boundaries} At 780\,nm the dispersive materials, including the absorbing silicon handle, reach the absorbing boundary. At 15 cells per wavelength a $1.5\times$ finer mesh moved the efficiency by 1.2\,dB, so the grating is designed at 30 cells per wavelength; at finer meshes the standard and the stable PML diverged. The adiabatic absorber does not diverge, but with its default 40 layers (about half a wavelength) it reflects a little, which adds lobes about 15\,nm apart to the spectrum. With the same grating and only the boundary changed (Fig.~\ref{fig:s-gcnum}b), the lobes appear only with the absorber, so they come from the boundary, not the grating. Averaged over one lobe period they disappear (Table~\ref{tab:s-boundary}), and the two boundaries then agree to 0.2\,dB at 780\,nm and the two finest absorber meshes to 0.01\,dB. The design uses the stable PML, whose spectrum is smooth.

\begin{table}[htbp]
\centering
\caption{The 780\,nm uniform grating with two absorbing boundaries; ``15\,nm mean'' is the average over 780$\pm$7.5\,nm, one lobe period.}\label{tab:s-boundary}
\begin{tabular}{lrrrr}
\toprule
Boundary & Cells per $\lambda$ & $\eta$(780\,nm) & 15\,nm mean & Ripple \\
\midrule
stable PML & 30 & 33.8\pct & 32.1\pct & $\pm2.4$\pct \\
absorber & 30 & 41.5\pct & 33.8\pct & $\pm7.3$\pct \\
absorber & 38 & 39.3\pct & 33.3\pct & $\pm6.3$\pct \\
absorber & 45 & 38.4\pct & 33.2\pct & $\pm6.2$\pct \\
\bottomrule
\end{tabular}
\end{table}

\paragraph{Apodisation at 780\,nm} Apodisation needs weak teeth at the start of the grating, so that the outgoing beam rises gently. TARANG measures the scattering strength $\alpha$ of uniform gratings from the decay of the guided field. At 780\,nm, fill factors 0.55--0.65 give $\alpha = 0.27$--0.32\,\textmu m$^{-1}$, decay lengths of 1.6--1.9\um, already shorter than the 2.5\um{} fibre waist. In a full etch the weaker teeth are the narrow trenches at fill factors of 0.70 and above, up to 0.79 with the 120\,nm trenches the process can clear; their simulations diverged for the boundary reason above. A partial etch would give weak teeth but adds a second lithography step. At 1550\,nm the apodised grating has trenches between 211 and 418\,nm, all wider than the 200\,nm limit set for that etch.

\paragraph{3D check of the focusing gratings} At 1550\,nm a 3D FDTD run of the focusing grating with apodised teeth, at 12 cells per wavelength, gives 28.0\pct{} (peak 28.1\pct{} at 1547\,nm), against 29.7\pct{} from the in-plane model. At 780\,nm the same check at 12 cells per wavelength (12 FlexCredits) would be dominated by mesh error, and a converged 3D mesh would cost about forty times more, so the 780\,nm focusing grating is quoted from the in-plane model (33.2\pct).

%% file: sections_supplementary/S5_grating_tolerances.tex
\section{Grating tolerances}\label{sec:s-gctol}

\begin{figure}[htbp]
\centering
\includegraphics[width=\textwidth]{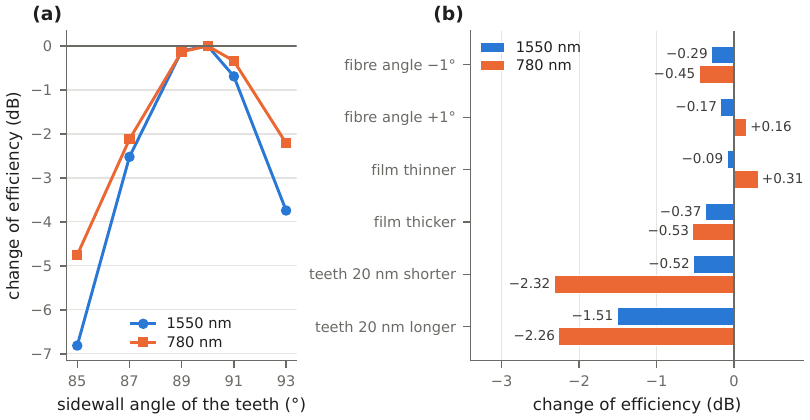}
\caption{Fabrication tolerance of the gratings at the design wavelength. (a) Sidewall angle of the teeth. (b) Fibre angle $\pm1^\circ$, film thickness ($\pm20$\,nm at 1550\,nm, $\pm10$\,nm at 780\,nm) and tooth length $\pm20$\,nm at fixed period.}\label{fig:s-gctol}
\end{figure}

Table~\ref{tab:s-gctol} and Fig.~\ref{fig:s-gctol} give the change of efficiency at the design wavelength for each fabrication error, at fixed period. The sidewall angle of the teeth dominates. At 780\,nm, tilting the fibre to 14$^\circ$ centres the peak (34.0\pct{} at 781\,nm).

\begin{table}[htbp]
\centering
\caption{Change of the grating efficiency at the design wavelength (dB).}\label{tab:s-gctol}
\begin{tabularx}{\textwidth}{Lll}
\toprule
Change & 1550\,nm & 780\,nm \\
\midrule
Fibre angle $-1^\circ$ / $+1^\circ$ & $-0.29$ / $-0.17$ & $-0.45$ / $+0.16$ \\
Film thinner / thicker ($\pm20$\,nm at 1550, $\pm10$\,nm at 780) & $-0.09$ / $-0.37$ & $+0.31$ / $-0.53$ \\
Teeth 20\,nm shorter / longer & $-0.52$ / $-1.51$ & $-2.32$ / $-2.26$ \\
Sidewall 89$^\circ$ / 87$^\circ$ / 85$^\circ$ & $-0.11$ / $-2.52$ / $-6.81$ & $-0.13$ / $-2.12$ / $-4.75$ \\
Sidewall 91$^\circ$ / 93$^\circ$ & $-0.69$ / $-3.74$ & $-0.34$ / $-2.21$ \\
\bottomrule
\end{tabularx}
\end{table}

\paragraph{Sensitivity to the period} The mask carries period variants of $\pm20$\,nm (1550\,nm) and $\pm10$\,nm (780\,nm). The two platforms respond differently. At a fill factor of 0.70 the 780\,nm peak moves about 2.4\,nm per 10\,nm of period (periods 0.562--0.579\um), against about 14\,nm expected from first-order phase matching, $\mathrm{d}\lambda/\mathrm{d}\Lambda = n_\mathrm{avg} - \sin\theta \approx \lambda/\Lambda$; at 1550\,nm it moves about 8\,nm per 10\,nm. The peak of the strongly reflecting 780\,nm grating is therefore not set by phase matching alone, and at 780\,nm the tooth-length variants, which move the peak by up to 9\,nm, tune the coupler more than the period variants do.

%% file: sections_supplementary/S6_whole_device.tex
\section{Whole-device model}\label{sec:s-device}

\begin{figure}[htbp]
\centering
\includegraphics[width=\textwidth]{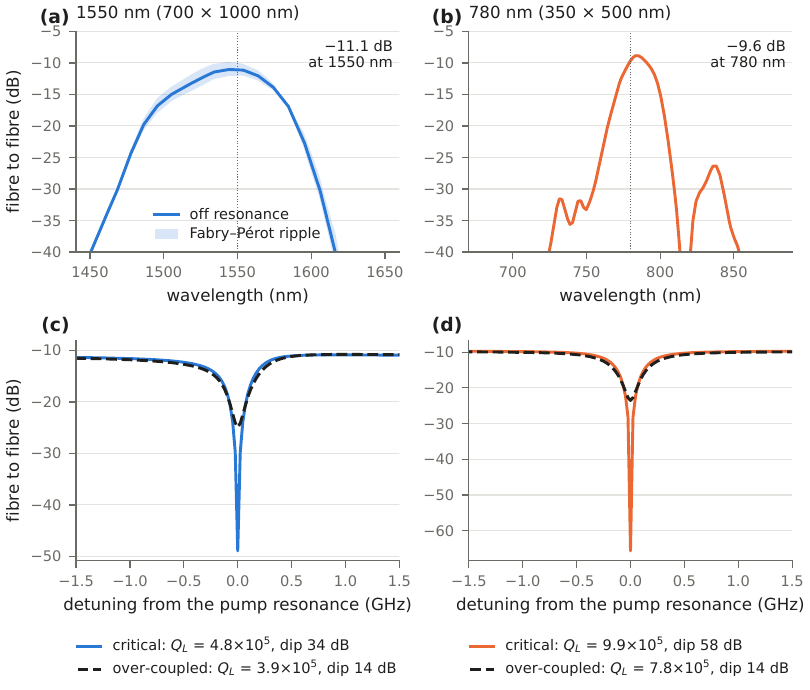}
\caption{Whole-device model, fibre to fibre. (a, b) Off-resonance transmission over the band (shaded: Fabry--P\'erot ripple between the two couplers, where the coupler reflection is known). (c, d) The pump resonance of the critically and over-coupled rings.}\label{fig:s-device}
\end{figure}

Stage 13 assembles the fibre-to-fibre transmission of the chip from the input coupler, the bus with the ring and the output coupler. Each ring resonance is placed at its frequency from $\Dint$ and given its own intrinsic and coupling $Q$; at 1550\,nm the reverse run of the apodised grating gives its reflection, so the Fabry--P\'erot ripple between the two couplers is included (2.0\,dB of ripple at the pump). At 780\,nm the checks stage did not include the reverse run, so the envelope is shown without ripple. Table~\ref{tab:s-budget} gives the loss budget and Fig.~\ref{fig:s-device} the spectra. At the pump the critically coupled rings have $Q_L = 4.8\times10^5$ (1550\,nm) and $9.9\times10^5$ (780\,nm) with dips of 34 and 58\,dB; the over-coupled rings $3.9\times10^5$ and $7.8\times10^5$ with dips of 14\,dB. With the converged 1550\,nm grating efficiency the off-resonance loss is about 1.1\,dB higher.

\begin{table}[htbp]
\centering
\caption{Fibre-to-fibre loss budget at the pump, off resonance (dB).}\label{tab:s-budget}
\begin{tabularx}{\textwidth}{Lrr}
\toprule
Item & 1550\,nm & 780\,nm \\
\midrule
Input coupler (1550: focusing grating, 3D FDTD; 780: uniform grating, 2D) & 5.53 & 4.72 \\
Output coupler (same design) & 5.53 & 4.72 \\
Focusing into the tip and tip tapers (1550: inside the 3D run; 780: in-plane and 2D models) & incl. & 0.15 \\
Propagation, about 410\um{} at 0.4\,dB/cm & 0.02 & 0.02 \\
\midrule
Fibre to fibre & 11.08 & 9.60 \\
\bottomrule
\end{tabularx}
\end{table}

%% file: sections_supplementary/S7_solitons.tex
\section{Soliton simulations}\label{sec:s-pylle}

TARANG writes the ring's integrated dispersion, in pyLLE's file format, and every ring parameter with units and definitions. These were handed unchanged to pyLLE~\cite{moille2019}, pumped at four times threshold. Table~\ref{tab:s-pylle} summarises the results. The simulations use $\pm110$ modes at 1550\,nm and $\pm200$ at 780\,nm, all inside the range of TARANG's $\Dint$, so pyLLE extrapolates nothing; at 1550\,nm a run with $\pm200$ modes, partly extrapolated, gives the same single soliton (127\,fs, 16.9\,mW). The $\Dint$ that pyLLE reconstructs from the exported file differs from TARANG's by at most 1.5\,MHz over 98\,GHz (1550\,nm) and 6.5\,MHz over 38\,GHz (780\,nm).

\begin{table}[htbp]
\centering
\caption{pyLLE simulations from the parameters TARANG exports.}\label{tab:s-pylle}
\begin{tabularx}{\textwidth}{Lll}
\toprule
 & 1550\,nm & 780\,nm \\
\midrule
Pump power for 4$\times$ threshold & 208\,mW & 25\,mW \\
Normalised pump $f^2$ & 3.64 & 3.60 \\
Soliton existence limit $\zeta_\mathrm{max} = \pi^2 f^2/8$~\cite{herr2014} & 4.49 & 4.45 \\
Detuning at which the sweep loses the soliton & 4.53 & 4.49 \\
Single soliton held at $\zeta$ & 3.11, one steady pulse & 3.09, one steady pulse \\
Comb power in the bus & 16.9\,mW & 0.78\,mW \\
Pulse FWHM, simulated / analytic~\cite{herr2014} & 130 / 144\,fs & 47 / 53\,fs \\
\bottomrule
\end{tabularx}
\end{table}

%% file: sections_supplementary/S8_sloped_walls.tex
\section{Published devices with sloped walls}\label{sec:s-sloped}

The largest differences in Table~3 are with Karpov et al.~\cite{karpov2018}, whose values are themselves simulations of a ring of radius about 23\um{} with a $1.45\times0.74$\um{} core and 80$^\circ$ sidewalls. The comparison in the main text uses vertical walls. Table~\ref{tab:s-sloped} repeats it with 80$^\circ$ walls, everything else unchanged: \SiO{} cladding (assumed; not stated in the paper), pump at 1060\,nm, the bent mode at $R = 23$\um, 70 mesh steps per wavelength with the two-sided average, and the paper's FSRs (1001\,GHz for \TE, 992\,GHz for \TM). The paper does not say at which plane of a sloped core the 1.45\um{} width holds, so it is solved with the width at the top, the middle and the bottom of the core.

\begin{table}[htbp]
\centering
\caption{$D_2/2\pi$ (MHz) of the Karpov et al. ring with vertical and 80$^\circ$ sidewalls.}\label{tab:s-sloped}
\begin{tabular}{llrr}
\toprule
Walls & 1.45\um{} width at & \TE{} & \TM{} \\
\midrule
90$^\circ$ & --- & $-38.0$ & 10.0 \\
80$^\circ$ & top (bottom 1.71\um) & $-57.3$ & $-1.7$ \\
80$^\circ$ & middle & $-38.2$ & 10.6 \\
80$^\circ$ & bottom (top 1.19\um) & $-17.7$ & 23.7 \\
\midrule
\multicolumn{2}{l}{Karpov et al.~\cite{karpov2018}, simulated} & $-32$ & 11 \\
\bottomrule
\end{tabular}
\end{table}

The plane of the width matters more than the wall angle itself. With the width at the middle of the core, 80$^\circ$ walls change $D_2$ by less than 1\pct{} (\TE) and 6\pct{} (\TM) from vertical walls. With the width at the top or the bottom they scale it by 0.47--1.51 (\TE) and 2.37 (\TM, width at the bottom), and for \TM{} with the width at the top the sign changes. Both published values lie inside the range spanned by the three planes, so the differences in Table~3 ($-18.8$\pct{} and $-8.9$\pct) are within the uncertainty of the published geometry.

%% file: sections_supplementary/S9_reproducing.tex
\section{Reproducing the results}\label{sec:s-repro}

Each design is defined by a short settings file (Listing~\ref{lst:s-toml}) and run in two steps. The command \texttt{python -m tarang.design -{}-config design.toml -{}-yes -{}-price-only} prices every cloud run and does the local work; the same command without \texttt{-{}-price-only} runs the design within the budget. It writes a design folder with the settings (\texttt{design.toml}); the results of every stage in machine-readable form; a \texttt{tables/} folder with every table as CSV; every figure with its data and settings; \texttt{report.md}, which lists every assumed value with its published source and the software versions, wall time and cost of each stage; the GDSII layout and parameter file; the Lugiato--Lefever inputs (\texttt{lle/}); and a benchmark pack for repeating the key simulations with another solver. Cloud results are cached by a fingerprint of the full simulation, so re-running a finished design reads every cloud result from the cache at no cost.

\begin{listing}[htbp]
\footnotesize
\begin{verbatim}
[run]
goal = "comb_ring"            # ring with bus and couplers
coupling = "grating"
wavelength_um = 1.55
max_cost_per_run = 6.0        # FlexCredits
max_total_cost = 40.0
[materials]
core = "si3n4"
box = "sio2"
cladding = "air"
[cross_section]
heights_um = [0.68, 0.7, 0.72]
widths_um = [0.6, 0.7, 0.8, 0.9, 1.0, 1.1, 1.2,
             1.3, 1.4, 1.5, 1.6, 1.7, 1.8]
mesh_ladder = [50, 60, 70, 80, 90]
tolerance_width_um = 0.05
tolerance_height_um = 0.02
[ring]
fsr_ghz = 200.0
loss_preset = "typical_ebl"   # 0.4 dB/cm, with its source
[grating]
etch = "full"
fiber_angle_deg = 12.0
mfd_um = 10.4
\end{verbatim}
\caption{Core settings of the 1550\,nm design (an excerpt of its \texttt{design.toml}). Every setting not given takes a documented default; the full file is in the design folder.}\label{lst:s-toml}
\end{listing}

\paragraph{Cost and run time} Table~\ref{tab:s-cost} gives the cloud cost and the local run time of the two designs. The cloud cost is TARANG's ledger of the estimates made before each run, summed over every simulation the design ran, including its checks. The local times are those of the final run, which solved every cross-section again with the two-sided average while all cloud results came from the cache. At 780\,nm the 2D grating runs at 30 cells per wavelength (153 runs, 39.8 FlexCredits) and the five 3D ring--bus runs (3.85 FlexCredits each) dominate, because halving the wavelength at the same ring radius makes the 3D coupler domain eight times larger in cells.

\begin{table}[htbp]
\centering
\caption{Cloud cost and local run time of the two designs (local times on the authors' Windows PC with four parallel solver processes).}\label{tab:s-cost}
\begin{tabularx}{\textwidth}{Lrr}
\toprule
 & 1550\,nm & 780\,nm \\
\midrule
Cloud simulations & 163 & 170 \\
Cloud cost (FlexCredits, estimates) & 12.0 & 62.6 \\
Local, cross-section stage (map, mesh ladder, tolerance box) & 68\,min & 466\,min \\
Local, ring stage (bend loss, integrated dispersion) & 11\,min & 52\,min \\
\bottomrule
\end{tabularx}
\end{table}

Every figure of the paper and of this document is drawn by a short Python script from the saved design folders only; no figure runs a simulation. The scripts, a snapshot of their input data and a README describing each figure's data and method will be deposited on Zenodo with the design data, linked from the software record~\cite{tarang_zenodo}. The results were computed with Python 3.11.16, TARANG v0.1.5, Tidy3D 2.12.0 and NumPy 2.4.6.